\documentclass[%
 reprint,
 amsmath,amssymb,
 aps,
]{revtex4-1}

\usepackage{graphicx}
\usepackage{dcolumn}
\usepackage{bm}

\begin{document}

\preprint{APS/123-QED}

\title{Functional integral for optical parametric amplification}

\author{Fuyong Wang}
 \altaffiliation{School of Physics, Shanghai Jiao Tong University, Shanghai, 200240, China.}





\begin{abstract}
It is demonstrated that the nature of optical parametric amplification is a quantum phenomenon. The system Lagrangian can be constructed by the path integral of coherent state. The equations of motion for photon operators are indeed the Euler-Lagrange equations of a Lagrangian. The quantum state evolution equation can also be obtained without resorting to quantum Hamiltonian or Lagrangian. Starting with classical Newton equation, quantum transition amplitude of the system can be educed by surface integral.
\end{abstract}

\maketitle


\section{Introduction}
Optical parametric amplification is an important physical phenomenon, in which occurs parametric frequency down conversion of light. Optical parametric fluorescence (also called spontaneous parametric down conversion) is the early stage of optical parametric amplification. In general, two models named signal model and idler model are coupled with pump light and amplified in the optical parametric amplification. Several theoretical studies\cite{Yariv,Gordon} have been done to predict this phenomenon before the observation of optical parametric fluorescence\cite{Harris,Doug}. The researches regarding optical parametric amplification are receiving renewed attention which is due to their wide-range utilization\cite{Mollo,Taps,Dorot,Alex,Thun}.

To explain where the signal and idler models originate from and how they get amplified, quantum theory must be adopted and classical analysis can only be applied to the amplification of the fields which already contain many quanta. To describe varying numbers of photons, second quantized operators named creation and annihilation operators are used to describe signal and idler models and the system Hamiltonian. Generally, the equations of motion for photon operators can be easily obtained with the help of Heisenberg equations of motion when the system Hamiltonian is given. Instead of using Heisenberg equations of motion, in this paper we deduce the system Lagrangian and the equations of motion for photon operators are obtained from Euler-Lagrange equation. The equations of motion obtained from commutation relations are consistent with those from Euler-Lagrangian equations. Besides, the propagetor of optical parametric amplification system can be figured out based on classical equations of motion and stringy quantization method. 

\section{Equations of motion}
To quantize electromagnetic field the vector potential is introduced and expressed as\cite{Rubin}\begin{align}
\begin{split}
\mathbf{A}(\mathbf{r},t)=\mathop{\sum}_{\mathbf{k}}\mathop{\sum}_{i=1}^2\sqrt{\frac{\hbar}{2\varepsilon_0V\omega_k}}\hat{\epsilon}_{\mathbf{k}i}[a_{\mathbf{k}i}(0)e^{i(\mathbf{k}\cdot\mathbf{r}-\omega_kt)}\\
+a_{\mathbf{k}i}^\dagger(0)e^{-i(\mathbf{k}\cdot\mathbf{r}-\omega_kt)}].
\end{split}
\end{align}
Here, $a$ and $a^\dagger$ are annihilation and creation operators. The field $\mathbf{A}$ is thus a collection of photons being created and destroyed each with energy $E_k=\hbar\omega_k$ and momentum $\mathbf{p}=\hbar\mathbf{k}$.
$\hat{\epsilon}_{\mathbf{k}i}$ describes the two possible and mutually orthogonal polarizations for the $\mathbf{k}th$ model.
For simplicity, transverse gauge is used. The electric field $\mathbf{E}$ and the magnetic field $\mathbf{B}$ can be expressed by $\mathbf{A}$\begin{equation}
\mathbf{E}=-\frac{\partial\mathbf{A}}{\partial t}, ~~\mathbf{B}=\nabla\times\mathbf{A}.
\end{equation}
We consider the following Hamitonian\cite{ROY}\begin{equation}
H=\frac{1}{2}\int(\varepsilon_0\mathbf{E}^2+\mu_0^{-1}\mathbf{B}^2)d\mathbf{r}.
\end{equation}
With above equations, the Hamiltonian of the free electromagnetic field is reduced to the infinite sum of Hamiltonians of independent harmonic oscillators\begin{equation}
H=\frac{1}{2}\mathop{\sum}_{\mathbf{k},i}\hbar\omega_k(a_{\mathbf{k}i}^\dagger a_{\mathbf{k}i}+a_{\mathbf{k}i}a_{\mathbf{k}i}^\dagger),
\end{equation}
where for each vector $\mathbf{k}$ there are two independent harmonic oscillators (for i=1,2). Considering the quantization conditions (commutation relations) $[a_k, a_{m}^\dagger]=\delta_{km}$, the Hamiltonian of the free electromagnetic field is reduced to the form\begin{equation}
H=\mathop{\sum}_{\mathbf{k},i}\hbar\omega_k(a_{\mathbf{k}i}^\dagger a_{\mathbf{k}i}+\frac{1}{2}).\label{001}
\end{equation}
Note that when applying quantization conditions, zero-point energy appears.

Quantum noise plays an important role in optical parametric amplification. The quantum uncertainty between electric and magnetic fields and the momentum (or velocity) fluctuations are two sources of quantum noise\cite{Charles}. The Hamiltonian of quantum noise can be expressed as Eq.\ref{001}. In addition, quantum noise is under thermal equilibrium. Therefore, according to the Boltzmann distribution law of statistical mechanics, the probability of finding a noise with energy $\hbar\omega_j$ is proportional to $exp(-\frac{\hbar\omega_j}{k_BT})$, where $k_B$ is Boltzmann's constant and $T$ is the system temperature.

For any given $\mathbf{k}$, the Hamiltonian is\begin{equation}
H_{\mathbf{k}}=\mathop{\sum}_{i=1}^2\hbar\omega_k(a_{i}^\dagger a_{i}+\frac{1}{2}).
\end{equation}
For simplicity, The pump photon takes one direction of polarization and its Hamiltonian is expressed as \begin{equation}
H_{p}=\hbar\omega_0(a_{0}^\dagger a_{0}+\frac{1}{2}).
\end{equation}

A coupling interaction between pump and quantum noise is provided by nonlinear medium and two models from quantum noise get coupled. These two models are named signal model and idler model, respectively, those frequencies add up to the frequency of the pump model. Ignoring zero-point energy, the Hamiltonian for optical parametric system is\begin{eqnarray}
H=H_0+H_{int},\label{0}
\end{eqnarray}
where\begin{equation}
H_0=\hbar\omega_0a_0^\dagger a_0+\hbar\omega_1a_1^\dagger a_1+\hbar\omega_2a_2^\dagger a_2,
\end{equation}
\begin{equation}
H_{int}=\hbar[\kappa a_0a_1^\dagger a_2^\dagger e^{-i\Delta \mathbf{k}\cdot \mathbf{r}}+\kappa^\dagger a_0^\dagger a_1 a_2 e^{i\Delta \mathbf{k}\cdot \mathbf{r}}]\label{1x}.
\end{equation}
The term $\kappa^\dagger a_0^\dagger a_1 a_2e^{i\Delta \mathbf{k}\cdot \mathbf{r}}$ in the Eq.\ref{1x} ensures $H_{int}$'s reality. Here, $\Delta \mathbf{k}$ represents phase mismatch, which satisfies $\Delta \mathbf{k}=\mathbf{k}_1+\mathbf{k}_2-\mathbf{k}_0$. The relation $\omega_0-(\omega_1+\omega_2)=0$ is used in the Eq.\ref{1x}.
Actually, the equations of motion for photon creation and annihilation operators can be easily obtained by using Heisenberg equations of motion if the system Hamiltonian is given. In this paper we deduce the system Lagrangian by the method in appendix.

The Lagrangian consists of two components $L=L_{free}+L_{int}$, where\begin{eqnarray}
L_{free}=\hbar[\frac{i}{2}(a_0^\dagger\dot{a}_0-\dot{a}_0^\dagger a_0)+\frac{i}{2}(a_1^\dagger\dot{a}_1-\dot{a}_1^\dagger a_1)\\\nonumber
+\frac{i}{2}(a_2^\dagger\dot{a}_2-\dot{a}_2^\dagger a_2)-\omega_0a_0^\dagger a_0-\omega_1a_1^\dagger a_1-\omega_2a_2^\dagger a_2],
\end{eqnarray}
\begin{equation}
L_{int}=-\hbar\kappa a_0a_1^\dagger a_2^\dagger e^{-i\Delta \mathbf{k}\cdot \mathbf{r}} -\hbar\kappa^\dagger a_0^\dagger a_1a_2 e^{i\Delta \mathbf{k}\cdot \mathbf{r}}.
\end{equation}

According to Euler-Lagrangian equation\begin{equation}
\frac{d}{dt}\frac{\partial L}{\partial \dot{a}^\dagger}-\frac{\partial L}{\partial a^\dagger}=0,
\end{equation}
we obtain\begin{eqnarray}
\frac{d a_0}{dt}=-i\omega_0a_0-i\kappa^\dagger a_1a_2e^{i\Delta \mathbf{k}\cdot \mathbf{r}},\\
\frac{d a_1}{dt}=-i\omega_1a_1-i\kappa a_0a_2^\dagger e^{-i\Delta \mathbf{k}\cdot \mathbf{r}},\\
\frac{d a_2}{dt}=-i\omega_2a_2-i\kappa a_0a_1^\dagger e^{-i\Delta \mathbf{k}\cdot \mathbf{r}}.
\end{eqnarray}
The above equations are actually quantum nonlinear coupled equations. The pump wave is usually very intense and we assume that the pump photon operator is just oscillating with time without depletion. That is to say, the pump photon operator approximately satisfies\begin{equation}
\frac{d a_0}{dt}=-i\omega_0a_0,
\end{equation}
and $a_0$ can be solved $a_0=a_0(0)e^{-i\omega_0t}$. Hence, the equations of signal and idler photon operators become\begin{align}
\begin{split}
\frac{d a_1}{dt}=-i\omega_1a_1-i\kappa a_0(0)e^{-i\omega_0t}a_2^\dagger e^{-i\Delta \mathbf{k}\cdot \mathbf{r}},\\
\frac{d a_2}{dt}=-i\omega_2a_2-i\kappa a_0(0)e^{-i\omega_0t}a_1^\dagger e^{-i\Delta \mathbf{k}\cdot \mathbf{r}}.\label{11}
\end{split}
\end{align}

\section{Path integral for parametric photons}
By solving Eq.\ref{11}, the equations of motion of photon operator are known. In addition, the evolution of the quantum states can be expressed as a functional integral. 
According to path integral theory, the transition amplitude of the parametric photons is\begin{eqnarray}\label{l1}
A(q_1,t_1|q_0,t_0)=\int[Dq ]exp\{ \frac{i}{\hbar}\mathop{\int}_{q(t)} L(q,t)dt\},
\end{eqnarray} 
where,$q=a_1a_2$.
It satisfies the evolution chain rule\begin{eqnarray}
A(q_1,t_1|q_2,t_0)=\int[Dq  ]A(q_1,t_1|q,t)A(q,t|q,t_0).
\end{eqnarray} 
Quantum states of the parametric photon system are described by the square integrable functions in the Hilbert space. Physical observables are Hermitian operators acting on such functions. The state $\varphi_1(q)$ at any later moment $t_1$ is able to be predicted when the state $\varphi_0(q)$ at initial moment $t_0$ is given according to the formula\begin{eqnarray}
\varphi_1(q)=\int[Dq^\prime]A(q,t_1|q^\prime ,t_0)\varphi_0(q^\prime).
\end{eqnarray} 

\section{Surface integral for parametric photons}

The system of optical parametric amplification is actually a dissipative system. Just as Brownian motion particles, parametric photons are modeled as particles undergoing a force provieded by pump. We start with classical equations to describe the two model parametric photons without resorting to Hamiltonian or Lagrangian.\begin{align}
\begin{split}
\ddot{x_1}=F_1,\\
\ddot{x_2}=F_2.\label{xcv}
\end{split}
\end{align} 
Like classical particles, parametric photons is treated as particles with unit mass, moving in one dimension under the action of the force $F$. There is a certain relationship between $F_1$ and $F_2$ for the parametric photons are twinship.

According to stringy quantization theory\cite{C13}, the transition amplitude can be expressed as a surface integral\begin{eqnarray}
A(x_1x_2,t_1|x_1x_2,t_0)\propto e^{\frac{i}{\hbar}S_{cl}}\int[D \Sigma  ]exp\{ \frac{i}{\hbar}\mathop{\int}_{\Sigma }\Omega \} ,
\end{eqnarray} 
where,$\Omega $ is a two-form\begin{equation}
  \Omega =\mathop{\sum }_{j=1,2} d(p_jdx_j-\frac{1}{2}p_j^2dt)+F_jdx_j\wedge dt.
\end{equation} 
The classical action $S_{cl}$ is the integral over classical curve. The surface integral can be converted to curve integral by using relation $$\mathop{\int}_{\Sigma}d\vartheta =\mathop{\int}_{\partial\Sigma }\vartheta $$. The transition amplitude becomes\begin{eqnarray}
A(x_1x_2,t_1|x_1x_2,t_0)\propto e^{\frac{i}{\hbar}S_{cl}}\int[D \Sigma  ]exp\{ \frac{i}{\hbar}\mathop{\int}_{\partial \Sigma }\vartheta \} ,
\end{eqnarray} 
where, \begin{align}
\begin{split}
\vartheta =\mathop{\sum }_{j=1,2}(p_jdx_j-\frac{1}{2}p_j^2dt)+F_jx_jdt\\
=\mathop{\sum }_{j=1,2}(p_j\dot{x_j} dt-\frac{1}{2}p_j^2dt)+F_jx_jdt
\end{split}
\end{align} 
The closed integral curve contains two paths with other two auxiliary curves canceling $$\mathop{\int}_{\partial \Sigma }=\mathop{\int}_{\gamma  }-\mathop{\int}_{\gamma _{cl}}$$.
We obtain\begin{eqnarray}
A(x_1x_2,t_1|x_1x_2,t_0)\propto \int[D \gamma]exp\{ \frac{i}{\hbar}\mathop{\int}_{\gamma }\vartheta \}.
\end{eqnarray}
The second quantization forms of canonical coordinate and momentum of harmonic oscillator are
$$x_j=\sqrt{\frac{\hbar}{2\omega_j}}(a_j+a_j^\dagger ),\\~~p_j=\sqrt{\frac{\hbar\omega_j}{2}}(a_j-a_j^\dagger ).$$
Adopting creation and annihilation operators, the transition amplitude is 
\begin{align}
\begin{split}
A(x_1x_2,t_1|x_1x_2,t_0)\propto\int[D\gamma]exp\{ \frac{i}{\hbar}\mathop{\int}_{\gamma  }\\
\mathop{\sum}_{j=1,2}[\hbar\frac{i}{2}(a_j^\dagger\dot{a}_j-\dot{a}_j^\dagger a_j) 
-\hbar\omega_ja_j^\dagger a_j+U(a_j)]\} .
\end{split}
\end{align}

The driving force $F(a_j)$ serves as a function of annihilating a pump photon to generate a signal photon and a idler photon. In order to simplify, $U(a_j)$ can be determined phenomenologically. The kernal of $U(a_j)$ should contain $a_0a_1^\dagger a_2^\dagger$, where $a_0$ is pump annihilation operator. In order to make $U(a_j)$ hermitian, a conjugate part $a_0^\dagger a_1a_2$ should be appear in $U(a_j)$. $U(a_j)$ is definitely has a form
\begin{equation}
U(a_j)=\hbar\eta a_0a_1^\dagger a_2^\dagger+\hbar\eta^\dagger a_0^\dagger a_1a_2,
\end{equation}
where, $\eta $ is a parameter determined by the coupling interaction of pump model with two parametric models.

Obviously, Eq.\ref{l1} is recovered.

\section{Conclusion}
Plenty of studies have been carried out to explain the quantum nature of optical parametric amplification. Although quantum noise plays an important role in optical parametric amplification, optical parametric amplification starts from zero-point energy. It seems that the lack of commutability among canonical coordinates and momenta leads to the emergence of zero-point energy. Whereas this non-commuting of canonical coordinates and momenta lies in the fact that wave function in Hilbert space is a function of space and time and the dynamical variables that can be measured are described by linear operators. 

We show the consistency between the Hamiltonian and the Lagrangian formalisms in the optical parametric amplification. The dynamical equations of photon operators obtained from Heisenberg equations of motion are the same as those from Euler-Lagrange equations.

Classical Newtonian mechanical equations seems more fundamental than quantum Lagrangian. Because quantum Lagrangian is always ambiguity. Starting with classical Newton equations, the dynamic equation of quantum states of optical parametric amplification system are worked out.  

\section{Appendix: Path integral for coherent states}

Coherent state $\mid\alpha>$ is an eigenstate of annihilation operator $a$\begin{equation}
a\mid\alpha>=\alpha\mid\alpha>, ~~~<\alpha\mid a^\dagger=<\alpha\mid\alpha^\ast.
\end{equation}
The motion of the state $\mid\alpha(t)>$ is determined by the system Hamiltonian which expressed by creation and annihilation operators.
The probability amplitude of the state propagating from $\mid\alpha(t_a)>$ to $\mid\alpha(t_b)>$ is \begin{equation}
K(\alpha(t_b),t_b;\alpha(t_a),t_a)=<\alpha(t_b)\mid U(t_b,t_a)\mid\alpha(t_a)>,
\end{equation}
which is also called propagator. Here, $U(t_b,t_a)$ is time development operator, which satisfies\begin{equation}
U(t_b,t_a)=e^{-\frac{i}{\hbar}\int_{t_{a}}^{t_{b}}Hdt}.
\end{equation}
We use path integral technic to calculate the propagator of coherent state with the Hamiltonian given in Eq.\ref{0}.\begin{eqnarray}\nonumber
&<&\alpha^0(t_b)\alpha^1(t_b)\alpha^2(t_b)\mid U(t_b,t_a)\mid\alpha^0(t_b)\alpha^1(t_a)\alpha^2(t_a)>\\\nonumber
&=&\int\cdots\int D_0D_1D_2\times<\alpha^0(t_b)\alpha^1(t_b)\alpha^2(t_b)\mid \\\nonumber
&&U(t_b,t_n)\mid\alpha^0(t_b)\alpha^1(t_n)\alpha^2(t_n)><\alpha^0(t_n)\alpha^1(t_n)\alpha^2(t_n)\mid \\\nonumber
&&U(t_n,t_{n-1})\mid\alpha^0(t_{n-1})\alpha^1(t_{n-1})\alpha^2(t_{n-1})><\alpha^0(t_{n-1})\\\nonumber
&&\alpha^1(t_{n-1})\alpha^2(t_{n-1})\mid\cdots<\alpha^0(t_{1})\alpha^1(t_{1})\alpha^2(t_{1})\mid \\
&&U(t_1,t_a)\mid\alpha^0(t_a)\alpha^1(t_a)\alpha^2(t_a)>,
\end{eqnarray}
where, $D_i=\pi^{-n}\Pi_{j=1}^{n}d^2\alpha_j^i$ and $i=0,1,2$. The relation $\int\frac{d^2\alpha}{\pi}\mid\alpha><\alpha\mid$ is used in above equation.

We first solve\begin{align}
\begin{split}
K_j=<\alpha^0_{j+1}\alpha^1_{j+1}\alpha^2_{j+1}\mid U(t_{j+1},t_j)\mid\alpha^0_{j}\alpha^1_{j}\alpha^2_j>=\\
<\alpha^0_{j+1}\alpha^1_{j+1}\alpha^2_{j+1}\mid e^{-\frac{i}{\hbar}\int_{t_{j}}^{t_{j+1}}Hdt}\mid\alpha^0_{j}\alpha^1_{j}\alpha^2_j>,
\end{split}
\end{align}
where $\alpha_{j}$ represents $\alpha(t_j)$. When $t_{j}$ is approaching to $t_{j+1}$ and to put it in another way $\eta=t_{j+1}-t_{j}$ is very small, the time development operator is approximately\begin{equation}
e^{-\frac{i}{\hbar}\int_{t_{j}}^{t_{j+1}}Hdt}\approx e^{-\frac{i}{\hbar}H\eta}\approx[1-\frac{i}{\hbar}H\eta].
\end{equation}
Therefore, with $\kappa^\prime=\kappa e^{-i\Delta \mathbf{k}}$, $K_j$ can be written as
\begin{eqnarray}\nonumber
K_j&=&<\alpha^0_{j+1}\alpha^1_{j+1}\alpha^2_{j+1}\mid [1-\frac{i}{\hbar}H\eta]\mid\alpha^0_{j}\alpha^1_{j}\alpha^2_j>\\\nonumber
&=&<\alpha^0_{j+1}\alpha^1_{j+1}\alpha^2_{j+1}\mid [1-i\eta\omega_0a^\dagger_0a_0-i\eta\omega_1a^\dagger_1a_1\\\nonumber
&-&i\eta\omega_2a^\dagger_2a_2-i\eta[\kappa^\prime a_0a^\dagger_{1}a^{\dagger}_2+\kappa^{\prime\dagger} a_0^\dagger a_{1}a_2]]\mid\alpha^0_{j}\alpha^1_{j}\alpha^2_j>\\\nonumber
&=&<\alpha^0_{j+1}\alpha^1_{j+1}\alpha^2_{j+1}\mid\alpha^0_{j}\alpha^1_{j}\alpha^2_j>\\\nonumber
&\times&[1-i\eta\omega_0\alpha_{j+1}^{0\ast}\alpha_j^0-i\eta\omega_1\alpha_{j+1}^{1\ast}\alpha_j^1-i\eta\omega_2\alpha_{j+1}^{2\ast}\alpha_j^2\\
&-&i\eta\kappa^\prime\alpha_{j+1}^{0}\alpha_{j+1}^{1\ast}\alpha_{j+1}^{2\ast}-i\eta\kappa^{\prime\dagger}\alpha_{j}^{0\ast}\alpha_{j}^{1}\alpha_{j}^{2}].
\end{eqnarray}

By using coherent states inner product relation $<\alpha\mid\alpha^\prime>=e^{-\mid\alpha\mid^2/2-\mid\alpha^\prime\mid^2/2+\alpha^\ast\alpha^\prime}$, we then find
\begin{eqnarray}\nonumber
K_j&=&exp(-\frac{1}{2}\mid\alpha_{j+1}^{0}\mid^2-\frac{1}{2}\mid\alpha_{j}^{0}\mid^2+\alpha_{j+1}^{0\ast}\alpha_{j}^{1}-\frac{1}{2}\\\nonumber
&&\mid\alpha_{j+1}^{1}\mid^2-\frac{1}{2}\mid\alpha_{1}^{1}\mid^2+\alpha_{j+1}^{1\ast}\alpha_{j}^{1}-\frac{1}{2}\\\nonumber
&&\mid\alpha_{j+1}^{2}\mid^2-\frac{1}{2}\mid\alpha_{j}^{2}\mid^2+\alpha_{j+1}^{2\ast}\alpha_{j}^{2})\times exp(-i\eta\omega_0\\\nonumber
&&\alpha_{j+1}^{0\ast}\alpha_{j}^{0}-i\eta\omega_1\alpha_{j+1}^{1\ast}\alpha_{j}^{1}-i\eta\omega_2\alpha_{j+1}^{2\ast}\alpha_{j}^{2}\\\nonumber
&+&\eta\kappa\alpha_{j+1}^{0}\alpha_{j+1}^{1\ast}\alpha_{j+1}^{2\ast}-\eta\kappa^\dagger\alpha_{j}^{0\ast}\alpha_{j}^{1}\alpha_{j}^{2})\\\nonumber
&=&exp(i\eta[\frac{i}{2}\alpha_{j+1}^{0\ast}\frac{\alpha_{j+1}^{0}-\alpha_{j}^{0}}{\eta}-\frac{i}{2}\frac{\alpha_{j+1}^{0\ast}-\alpha_{j}^{0\ast}}{\eta}\alpha_{j}^{0}\\\nonumber
&+&\frac{i}{2}\alpha_{j+1}^{1\ast}\frac{\alpha_{j+1}^{1}-\alpha_{j}^{1}}{\eta}-\frac{i}{2}\frac{\alpha_{j+1}^{1\ast}-\alpha_{j}^{1\ast}}{\eta}\alpha_{j}^{1}+\frac{i}{2}\alpha_{j+1}^{2\ast}\\\nonumber
&&\frac{\alpha_{j+1}^{2}-\alpha_{j}^{2}}{\eta}-\frac{i}{2}\frac{\alpha_{j+1}^{2\ast}-\alpha_{j}^{2\ast}}{\eta}\alpha_{j}^{2}-\omega_0\alpha_{j+1}^{0\ast}\alpha_{j}^{0}-\omega_1\alpha_{j+1}^{1\ast}\\
&&\alpha_{j}^{1}-\omega_2\alpha_{j+1}^{2\ast}-\kappa^\prime\alpha_{j+1}^{0}\alpha_{j+1}^{1\ast}\alpha_{j+1}^{2\ast}-\kappa^{\prime\dagger}\alpha_{j}^{0\ast}\alpha_{j}^{1}\alpha_{j}^{2}].
\end{eqnarray}

When $n\rightarrow\infty$, the time interval $\eta\rightarrow 0$, the final expression of the propagator is\begin{eqnarray}\nonumber
&&<\alpha^0(t_b)\alpha^1(t_b)\alpha^2(t_b)\mid U(t_b,t_a)\mid\alpha^0(t_a)\alpha^1(t_a)\alpha^2(t_a)>\\\nonumber
&=&\int\cdots\int D_0D_1D_2\times exp(\frac{i}{\hbar}\int_{t_a}^{t_b}dt\{\hbar[\frac{i}{2}(\alpha_{0}^{\ast}\dot{\alpha}_0-\dot{\alpha}_0^\ast\alpha_0)\\\nonumber
&+&\frac{i}{2}(\alpha_{1}^{\ast}\dot{\alpha}_1-\dot{\alpha}_1^\ast\alpha_1)+\frac{i}{2}(\alpha_{2}^{\ast}\dot{\alpha}_2-\dot{\alpha}_2^\ast\alpha_2)-\omega_0\alpha_0^\ast\alpha_0\\
&-&\omega_1\alpha_1^\ast\alpha_1-\hbar\omega_2\alpha_2^\ast\alpha_2-\kappa^\prime\alpha_0\alpha_{1}^{\ast}\alpha_{2}^{\ast}-\kappa^{\prime\dagger}\alpha_0^\ast\alpha_{1}\alpha_{2}]\}).
\end{eqnarray}

The exponential part of the above equation is the time integral of the Lagrangian $\int_{t_a}^{t_b}L(t)dt$. Therefore, the corresponding Lagrangian is\begin{align}
\begin{split}
L=\hbar[\frac{i}{2}(a_0^\dagger\dot{a}_0-\dot{a}_0^\dagger a_0)+\frac{i}{2}(a_1^\dagger\dot{a}_1-\dot{a}_1^\dagger a_1)+\frac{i}{2}(a_2^\dagger\dot{a}_2-\dot{a}_2^\dagger a_2)\\
-\omega_0a_0^\dagger a_0-\omega_1a_1^\dagger a_1-\omega_2a_2^\dagger a_2-\kappa^\prime a_0a_1^\dagger a_2^\dagger -\kappa^{\prime\dagger} a_0^\dagger a_1a_2].
\end{split}
\end{align}




\end{document}